%Paper: hep-th/9312201
%From: SEZGIN@PHYS.TAMU.EDU
%Date: Tue, 28 Dec 1993 15:24:05 CST

%%%%%%%%%%%%%%%%%%%%%%%%%%%%%%%%%%%%%%%%%%%%%%%%%%%%%%%%%%%%%%%
%%%                                                         %%%
%%%                                                         %%%
%%%                                                         %%%
%%%   On Nonlinear Superconformal Algebras With $N > 4$     %%%
%%%                                                         %%%
%%%                                                         %%%
%%%                                                         %%%
%%%                   Z. Khviengia and E. Sezgin            %%%
%%%                                                         %%%
%%%                          USE PLAIN TEX                  %%%
%%%                                                         %%%
%%%%%%%%%%%%%%%%%%%%%%%%%%%%%%%%%%%%%%%%%%%%%%%%%%%%%%%%%%%%%%%

%manumac.tex version 3.0%
\catcode`@=11

\def\singlespace{\normalbaselines}
\def\oneandahalfspace{\baselineskip=1.15\normalbaselineskip plus 1pt
\lineskip=2pt\lineskiplimit=1pt}

\def\np{\vfill\eject}

\def\nofirstpagenoten{\nopagenumbers\footline={\ifnum\pageno>1\tenrm
\hss\folio\hss\fi}}
\def\nofirstpagenotwelve{\nopagenumbers\footline={\ifnum\pageno>1\twelverm
\hss\folio\hss\fi}}
\def\leaderfill{\leaders\hbox to 1em{\hss.\hss}\hfill}
\def\ft#1#2{{\textstyle{{#1}\over{#2}}}}
\def\frac#1/#2{\leavevmode\kern.1em
\raise.5ex\hbox{\the\scriptfont0 #1}\kern-.1em/\kern-.15em
\lower.25ex\hbox{\the\scriptfont0 #2}}
\def\sfrac#1/#2{\leavevmode\kern.1em
\raise.5ex\hbox{\the\scriptscriptfont0 #1}\kern-.1em/\kern-.15em
\lower.25ex\hbox{\the\scriptscriptfont0 #2}}

\parindent=20pt
\def\narrow{\advance\leftskip by 40pt \advance\rightskip by 40pt}

\def\AB{\bigskip
        \centerline{\bf ABSTRACT}\medskip\narrow}
\def\nonarrower{\advance\leftskip by -40pt\advance\rightskip by -40pt}
\def\AE{\bigskip\nonarrower}

\def\boxit#1{\vbox{\hrule\hbox{\vrule\kern3pt
        \vbox{\kern3pt#1\kern3pt}\kern3pt\vrule}\hrule}}

\def\gtorder{\mathrel{\raise.3ex\hbox{$>$}\mkern-14mu
             \lower0.6ex\hbox{$\sim$}}}
\def\ltorder{\mathrel{\raise.3ex\hbox{$<$}|mkern-14mu
             \lower0.6ex\hbox{\sim$}}}
\def\dalemb#1#2{{\vbox{\hrule height .#2pt
        \hbox{\vrule width.#2pt height#1pt \kern#1pt
                \vrule width.#2pt}
        \hrule height.#2pt}}}
\def\square{\mathord{\dalemb{4.9}{5}\hbox{\hskip1pt}}}

\font\fourteentt=cmtt10 scaled \magstep2
\font\fourteenbf=cmbx12 scaled \magstep1
\font\fourteenrm=cmr12 scaled \magstep1
\font\fourteeni=cmmi12 scaled \magstep1
\font\fourteenssr=cmss12 scaled \magstep1
\font\fourteenmbi=cmmib10 scaled \magstep2
\font\fourteensy=cmsy10 scaled \magstep2
\font\fourteensl=cmsl12 scaled \magstep1
\font\fourteenex=cmex10 scaled \magstep2
\font\fourteenit=cmti12 scaled \magstep1
\font\twelvett=cmtt12 \font\twelvebf=cmbx12
\font\twelverm=cmr12  \font\twelvei=cmmi12
\font\twelvessr=cmss12 \font\twelvembi=cmmib10 scaled \magstep1
\font\twelvesy=cmsy10 scaled \magstep1
\font\twelvesl=cmsl12 \font\twelveex=cmex10 scaled \magstep1
\font\twelveit=cmti12
\font\tenssr=cmss10 \font\tenmbi=cmmib10
 
 \font\ninebf=cmbx9
\font\ninerm=cmr9  \font\ninei=cmmi9
\font\ninesy=cmsy9 \font\ninessr=cmss9
\font\ninembi=cmmib10 scaled 900
\font\eightit=cmti8 \font\eightsl=cmsl8
\font\eighttt=cmtt8 \font\eightbf=cmbx8
\font\eightrm=cmr8  \font\eighti=cmmi8
\font\eightsy=cmsy8 \font\eightex=cmex10 scaled 800
\font\eightssr=cmss8 \font\eightmbi=cmmib10 scaled 800
 
\font\sevenbf=cmbx7 \font\sevenrm=cmr7 \font\seveni=cmmi7
\font\sevensy=cmsy7 
\font\sevenssr=cmss9 scaled 778 \font\sevenmbi=cmmib10 scaled 700
 
 \font\sixbf=cmbx7 scaled 875
\font\sixrm=cmr6  \font\sixi=cmmi6
\font\sixsy=cmsy6 \font\sixssr=cmss8 scaled 750
\font\sixmbi=cmmib10 scaled 600
\font\fivessr=cmss8 scaled 625  \font\fivembi=cmmib10 scaled 500

\newskip\ttglue
\newfam\ssrfam
\newfam\mbifam

\mathchardef\alpha="710B
\mathchardef\beta="710C
\mathchardef\gamma="710D
\mathchardef\delta="710E
\mathchardef\epsilon="710F
\mathchardef\zeta="7110
\mathchardef\eta="7111
\mathchardef\theta="7112
\mathchardef\iota="7113
\mathchardef\kappa="7114
\mathchardef\lambda="7115
\mathchardef\mu="7116
\mathchardef\nu="7117
\mathchardef\xi="7118
\mathchardef\pi="7119
\mathchardef\rho="711A
\mathchardef\sigma="711B
\mathchardef\tau="711C
\mathchardef\upsilon="711D
\mathchardef\phi="711E
\mathchardef\chi="711F
\mathchardef\psi="7120
\mathchardef\omega="7121
\mathchardef\varepsilon="7122
\mathchardef\vartheta="7123
\mathchardef\varpi="7124
\mathchardef\varrho="7125
\mathchardef\varsigma="7126
\mathchardef\varphi="7127
\mathchardef\partial="7140

\def\fourteenpoint{\def\rm{\fam0\fourteenrm}
\textfont0=\fourteenrm \scriptfont0=\tenrm \scriptscriptfont0=\sevenrm
\textfont1=\fourteeni \scriptfont1=\teni \scriptscriptfont1=\seveni
\textfont2=\fourteensy \scriptfont2=\tensy \scriptscriptfont2=\sevensy
\textfont3=\fourteenex \scriptfont3=\fourteenex \scriptscriptfont3=\fourteenex
\def\it{\fam\itfam\fourteenit} \textfont\itfam=\fourteenit
\def\sl{\fam\slfam\fourteensl} \textfont\slfam=\fourteensl
\def\bf{\fam\bffam\fourteenbf} \textfont\bffam=\fourteenbf
\scriptfont\bffam=\tenbf \scriptscriptfont\bffam=\sevenbf
\def\tt{\fam\ttfam\fourteentt} \textfont\ttfam=\fourteentt
\def\ssr{\fam\ssrfam\fourteenssr} \textfont\ssrfam=\fourteenssr
\scriptfont\ssrfam=\tenmbi \scriptscriptfont\ssrfam=\sevenmbi
\def\mbi{\fam\mbifam\fourteenmbi} \textfont\mbifam=\fourteenmbi
\scriptfont\mbifam=\tenmbi \scriptscriptfont\mbifam=\sevenmbi
\tt \ttglue=.5em plus .25em minus .15em
\normalbaselineskip=16pt
\bigskipamount=16pt plus5pt minus5pt
\medskipamount=8pt plus3pt minus3pt
\smallskipamount=4pt plus1pt minus1pt
\abovedisplayskip=16pt plus 4pt minus 12pt
\belowdisplayskip=16pt plus 4pt minus 12pt
\abovedisplayshortskip=0pt plus 4pt
\belowdisplayshortskip=9pt plus 4pt minus 6pt
\parskip=5pt plus 1.5pt
\twelvefoot
\setbox\strutbox=\hbox{\vrule height12pt depth5pt width0pt}
\let\sc=\tenrm
\let\big=\fourteenbig \normalbaselines\rm}
\def\fourteenbig#1{{\hbox{$\left#1\vbox to12pt{}\right.\n@space$}}
\def\square{\mathord{\dalemb{6.8}{7}\hbox{\hskip1pt}}}}

\def\twelvepoint{\def\rm{\fam0\twelverm}
\textfont0=\twelverm \scriptfont0=\ninerm \scriptscriptfont0=\sevenrm
\textfont1=\twelvei \scriptfont1=\ninei \scriptscriptfont1=\seveni
\textfont2=\twelvesy \scriptfont2=\ninesy \scriptscriptfont2=\sevensy
\textfont3=\twelveex \scriptfont3=\twelveex \scriptscriptfont3=\twelveex
\def\it{\fam\itfam\twelveit} \textfont\itfam=\twelveit
\def\sl{\fam\slfam\twelvesl} \textfont\slfam=\twelvesl
\def\bf{\fam\bffam\twelvebf} \textfont\bffam=\twelvebf
\scriptfont\bffam=\ninebf \scriptscriptfont\bffam=\sevenbf
\def\tt{\fam\ttfam\twelvett} \textfont\ttfam=\twelvett
\def\ssr{\fam\ssrfam\twelvessr} \textfont\ssrfam=\twelvessr
\scriptfont\ssrfam=\ninessr \scriptscriptfont\ssrfam=\sevenssr
\def\mbi{\fam\mbifam\twelvembi} \textfont\mbifam=\twelvembi
\scriptfont\mbifam=\ninembi \scriptscriptfont\mbifam=\sevenmbi
\tt \ttglue=.5em plus .25em minus .15em
\normalbaselineskip=14pt
\bigskipamount=14pt plus4pt minus4pt
\medskipamount=7pt plus2pt minus2pt
\abovedisplayskip=14pt plus 3pt minus 10pt
\belowdisplayskip=14pt plus 3pt minus 10pt
\abovedisplayshortskip=0pt plus 3pt
\belowdisplayshortskip=8pt plus 3pt minus 5pt
\parskip=3pt plus 1.5pt
\tenfoot
\setbox\strutbox=\hbox{\vrule height10pt depth4pt width0pt}
\let\sc=\ninerm
\let\big=\twelvebig \normalbaselines\rm}
\def\twelvebig#1{{\hbox{$\left#1\vbox to10pt{}\right.\n@space$}}
\def\square{\mathord{\dalemb{5.9}{6}\hbox{\hskip1pt}}}}

\def\tenpoint{\def\rm{\fam0\tenrm}
\textfont0=\tenrm \scriptfont0=\sevenrm \scriptscriptfont0=\fiverm
\textfont1=\teni \scriptfont1=\seveni \scriptscriptfont1=\fivei
\textfont2=\tensy \scriptfont2=\sevensy \scriptscriptfont2=\fivesy
\textfont3=\tenex \scriptfont3=\tenex \scriptscriptfont3=\tenex
\def\it{\fam\itfam\tenit} \textfont\itfam=\tenit
\def\sl{\fam\slfam\tensl} \textfont\slfam=\tensl
\def\bf{\fam\bffam\tenbf} \textfont\bffam=\tenbf
\scriptfont\bffam=\sevenbf \scriptscriptfont\bffam=\fivebf
\def\tt{\fam\ttfam\tentt} \textfont\ttfam=\tentt
\def\ssr{\fam\ssrfam\tenssr} \textfont\ssrfam=\tenssr
\scriptfont\ssrfam=\sevenssr \scriptscriptfont\ssrfam=\fivessr
\def\mbi{\fam\mbifam\tenmbi} \textfont\mbifam=\tenmbi
\scriptfont\mbifam=\sevenmbi \scriptscriptfont\mbifam=\fivembi
\tt \ttglue=.5em plus .25em minus .15em
\normalbaselineskip=12pt
\bigskipamount=12pt plus4pt minus4pt
\medskipamount=6pt plus2pt minus2pt
\abovedisplayskip=12pt plus 3pt minus 9pt
\belowdisplayskip=12pt plus 3pt minus 9pt
\abovedisplayshortskip=0pt plus 3pt
\belowdisplayshortskip=7pt plus 3pt minus 4pt
\parskip=0.0pt plus 1.0pt
\eightfoot
\setbox\strutbox=\hbox{\vrule height8.5pt depth3.5pt width0pt}
\let\sc=\eightrm
\let\big=\tenbig \normalbaselines\rm}
\def\tenbig#1{{\hbox{$\left#1\vbox to8.5pt{}\right.\n@space$}}
\def\square{\mathord{\dalemb{4.9}{5}\hbox{\hskip1pt}}}}

\def\eightpoint{\def\rm{\fam0\eightrm}
\textfont0=\eightrm \scriptfont0=\sixrm \scriptscriptfont0=\fiverm
\textfont1=\eighti \scriptfont1=\sixi \scriptscriptfont1=\fivei
\textfont2=\eightsy \scriptfont2=\sixsy \scriptscriptfont2=\fivesy
\textfont3=\eightex \scriptfont3=\eightex \scriptscriptfont3=\eightex
\def\it{\fam\itfam\eightit} \textfont\itfam=\eightit
\def\sl{\fam\slfam\eightsl} \textfont\slfam=\eightsl
\def\bf{\fam\bffam\eightbf} \textfont\bffam=\eightbf
\scriptfont\bffam=\sixbf \scriptscriptfont\bffam=\fivebf
\def\tt{\fam\ttfam\eighttt} \textfont\ttfam=\eighttt
\def\ssr{\fam\ssrfam\eightssr} \textfont\ssrfam=\eightssr
\scriptfont\ssrfam=\sixssr \scriptscriptfont\ssrfam=\fivessr
\def\mbi{\fam\mbifam\eightmbi} \textfont\mbifam=\eightmbi
\scriptfont\mbifam=\sixmbi \scriptscriptfont\mbifam=\fivembi
\tt \ttglue=.5em plus .25em minus .15em
\normalbaselineskip=9pt
\bigskipamount=9pt plus3pt minus3pt
\medskipamount=5pt plus2pt minus2pt
\abovedisplayskip=9pt plus 3pt minus 9pt
\belowdisplayskip=9pt plus 3pt minus 9pt
\abovedisplayshortskip=0pt plus 3pt
\belowdisplayshortskip=5pt plus 3pt minus 4pt
\parskip=0.0pt plus 1.0pt
\setbox\strutbox=\hbox{\vrule height8.5pt depth3.5pt width0pt}
\let\sc=\sixrm
\let\big=\eightbig \normalbaselines\rm}
\def\eightbig#1{{\hbox{$\left#1\vbox to6.5pt{}\right.\n@space$}}
\def\square{\mathord{\dalemb{3.9}{4}\hbox{\hskip1pt}}}}

\def\vfootnote#1{\insert\footins\bgroup\footsuite
    \interlinepenalty=\interfootnotelinepenalty
    \splittopskip=\ht\strutbox
    \splitmaxdepth=\dp\strutbox \floatingpenalty=20000
    \leftskip=0pt \rightskip=0pt \spaceskip=0pt \xspaceskip=0pt
    \textindent{#1}\footstrut\futurelet\next\fo@t}
\def\hangfootnote#1{\edef\@sf{\spacefactor\the\spacefactor}#1\@sf
    \insert\footins\bgroup\footsuite
    \let\par=\endgraf
    \interlinepenalty=\interfootnotelinepenalty
    \splittopskip=\ht\strutbox
    \splitmaxdepth=\dp\strutbox \floatingpenalty=20000
    \leftskip=0pt \rightskip=0pt \spaceskip=0pt \xspaceskip=0pt
    \smallskip\item{#1}\bgroup\strut\aftergroup\@foot\let\next}
\def\footsuite{}
\def\twelvefoot{\def\footsuite{\twelvepoint}}
\def\tenfoot{\def\footsuite{\tenpoint}}
\def\eightfoot{\def\footsuite{\eightpoint}}
\catcode`@=12

\oneandahalfspace
\twelvepoint
\rightline{CTP TAMU--71/93}
\rightline{hep-th/9312201}
\rightline{December 1993}

\vskip 2truecm
\centerline{{\bf On Nonlinear Superconformal Algebras
With $N > 4$}\hangfootnote{$^*$}{\tenfoot Contribution to the Proceedings
of the Trieste Summer School in High Energy Physics and Cosmology, \ \ \ \ \ \
\ \  14 June-30 July 1993, Trieste, Italy.}}
 \vskip 1.5truecm \centerline{Z. Khviengia and E.
Sezgin\footnote{$^\dagger$}{\tenfoot \sl  Supported in part by the National
Science Foundation, under grant PHY-9106593.}}

\vskip 1.5truecm
\centerline{\it  Center
for Theoretical Physics,
Texas A\&M University,}
\centerline{\it College Station, TX 77843--4242, USA.}
\vskip 1.5truecm
\AB\singlespace
   {\tenfoot We discuss the structure, realizations and quantum BRST
operators of a class of nonlinear superconformal algebras with $N>4$.}
 \AE\oneandahalfspace
\np
\vskip 1.5truecm
\noindent{\bf 1. Introduction}
\bigskip
A class of superconformal algebras with $N$-extended superymmetry were
classified long ago [1]. They contain a single spin 2 generator and multiplets
of other generators of
spin decreasing by half  units down to minimum spin $2-{1\over 2}N$.
These are linear algebras and the $N=1,2,4$ cases have found important and
interesting applications in string
theory, but beyond $N=4$, concordant with the fact that negative dimension
generators set in, no such applications have emerged so far.

One way to extend the supersymmetry beyond $N=4$ is to allow nonlinearities
in the algebra. The mildest such nonlinearity would be quadratic.
Requring one spin 2 generator, multiplets of spin 3/2 and spin 1
generators, generalizations involving quadratic nonlinearity have been found
[2,3]. They contain the energy momentum tensor, spin 3/2 currents
in the fundamental representation of $so_N$ or $u_N$, and spin 1
currents in the adjoint representation of the same algebras. A
characteristic feature of these algebras is that the  OPE of two spin 3/2
currents contains an operator bilinear in spin 1 currents.

In general, a convenient way of characterizing the quadratically nonlinear
algebras of the above type is to specify a pair $(g,\rho)$ where $g$ is the
(super) Lie algebra and $\rho$ is the representation carried by the spin 3/2
currents. All possibilities for this pair (as determined by the closure
of the algebra) have been classified in [4-8], including the cases when the
spin 3/2 generators are {\it commuting} (corresponding to the
quasi-superconformal algebras) and the doubly graded superalgebras where the
affine Lie algebra sector itself is based on a superalgebra. The
(quasi) superconformal algebras with simple $g$ and irreducible $\rho$ are
listed below.

%%%%%%%%%%%%%%%%%%%% THE TABLE HERE %%%%%%%%%%%%%%%%%%%%%%%%%%

\bigskip
\vbox{\tabskip=0pt\offinterlineskip
\def\tablerule{\noalign{\hrule}}
\halign to 300pt{\strut#&\vrule#\tabskip=0em plus2.5em&
\hfil#&\vrule#&\hfil#&\vrule#&\hfil#&\vrule#&
\hfil#&\vrule#&\hfil#&\vrule#
\tabskip=0pt\cr\tablerule
&&\omit\hidewidth $g$ \hidewidth
&&\omit\hidewidth $\rho$ \hidewidth
&&\omit\hidewidth $h_g^\vee$ \hidewidth
&&\omit\hidewidth $i_\rho$\hidewidth
&&\omit \hidewidth$\psi^2$\hidewidth&\cr\tablerule
&&\multispan9\hfil superconformal\quad ($\epsilon=-1$)\hfil&\cr\tablerule
&&$D_n$&&$2n$&&$2n-2$ &&1&&2\qquad&\cr\tablerule
&&$B_n$&&$2n+1$&&$2n-1$ &&1&&4\qquad&\cr\tablerule
&&$B_3$&&$8_s$&&5 &&1&&4\qquad&\cr\tablerule
&&$G_2$&&7&&4 &&1&&6\qquad&\cr\tablerule
&&\multispan9\hfil
quasisuperconformal\quad ($\epsilon=1$)\hfil&\cr\tablerule
&&$C_n$&&$2n$&&$n+1$ &&$1/2$&&4\qquad&\cr\tablerule
&&$A_5$&&20&&6 &&3&&2\qquad&\cr\tablerule
&&$D_6$&&32&&10 &&4&&2\qquad&\cr\tablerule
&&$E_7$&&56&&18 &&6&&2\qquad&\cr\tablerule
&&$C_3$&&14&&4 &&$5/2$&&4\qquad&\cr\tablerule
&&$A_1$&&4&&2 &&$5/2$&&2\qquad&\cr\tablerule}}
\bigskip

%%%%%%%%%%%%%%%%%%%%%%%%%%%%%%%%%%%%%%%%%%%%%%%%%%%%%%%%%%

Our main motivation
for studying these algebras is the possibility of their use in
constructing
a novel string theory with $N>4$ worldsheet {\it local}  supersymmetry.
Our recent work [9] on quadratically nonlinear superconformal algebras dealt
with the construction of the quantum BRST operator, which is a step towards
this goal. It generalizes the work of ref. [10] where an elegant formula was
derived for the classical and quantum BRST operator for a large class of
quadratically nonlinear algebras. In section 2, we shall summarize the
results of refs. [9,10]. The issue of free field realization is
another important aspect of string theory construction. In section 3 we
summarize result of ref. [11] which furnishes the most general realization
known so far. Quadratic nonlinearity is the mildest one that can be
introduced. Another kind of highly nonlinear $N=8$ superconformal algebra,
based on the seven  sphere, has also been considered [12]. We shall briefly
discuss the result of [12] on this algebra in section 4.
\bigskip
\noindent{\bf 2. BRST Operator for the Quadratically Nonlinear
                 Superconformal Algebra}
\bigskip
The algebras of the type listed in the Table  are
generated by the
energy-momentum tensor $T(z)$, the dimension 3/2 supercurrents $G^i(z),
i=1,...,{\rm dim}\ \rho := d$ and the dimension 1 currents $J^a(z),
a=1,...,{\rm dim}\ g := D$. The products
$T(z) T(\omega)$, $T(z)G^i(\omega)$ and $T(z)J^a(\omega)$ are standard. The
remaning part of the operator product algebra takes the form
$$
\eqalignno{
G^i(z)G^j(\omega) =& {b\eta^{ij}\over
(z-\omega)^3}+ {\sigma \lambda^{ij}_a J^a(\omega)\over (z-\omega)^2} +{\ft12
\sigma \lambda^{ij}_a \partial J^a(\omega)\over (z-\omega)} +{2\eta^{ij}
T(\omega)\over (z-\omega)} \cr
& +{\gamma P^{ij}_{ab} (J^a J^b)(\omega)\over (z-\omega)}+\cdots \ ,&(2.1a)\cr
J^a(z) G^i(\omega) =& {-\lambda^{ai}_{~~j} G^j(\omega)\over (z-\omega)}+\cdots
\ , &(2.1b)\cr
 J^a(z) J^b(\omega) =& {-\ft12 k\psi^2 \delta^{ab}\over (z-\omega)^2}+
{f^{ab}_{~~~c} J^c(\omega)\over (z-\omega)}+\cdots \ ,&(2.1c)\cr}
$$
where the generators $\lambda^a_{ij}$  and the structure constants
$f_{ab}^{~~c}$ satisfy the relations [4]
$$
\eqalign{
\lambda_{aik}\lambda^k_{bj}-
\lambda_{bik}\lambda^k_{aj}
 = f_{ab}^{~~c} \lambda_{ci}^{~~j} \ ,  \quad\quad
\lambda_{aij}\lambda_b^{ji}=-i_\rho \psi^2 \delta_{ab}  \ , \cr
(\lambda^a_{ij}\lambda^{a}_{k\ell}-\lambda^a_{jk}\lambda^{a}_{i\ell})=
{2\epsilon\over
\sigma_0} (\eta_{ij}\eta_{kl}+\eta_{jk}\eta_{i\ell}-2\eta_{ki}\eta_{j\ell}),
\qquad  \sigma_0=2(d+\epsilon)/C_\rho\ ,  \cr}\eqno(2.2)
$$
and the quadratic nonlinearity is defined by  $(JJ)(\omega) :=
{1\over 2\pi i}\oint d\zeta {J(\zeta)J(\omega)\over (\zeta-\omega)}$.
The Cartan-Killing metric $g^{ab}$ is defined as: $
g_{ab}=f_{ac}^{~~d}f_{bd}^{~~c} =-C_v \delta_{ab}$. The Lie algebra is taken
to be complex for the time being.  The Dynkin index $i_\rho$ of the
representation $\rho$ is defined by $i_\rho={dC_\rho\over D\psi^2}$, where
$C_\rho$ is the eigenvalue of the second Casimir in the representation $\rho$
defined by  $ \lambda^a_{ik}\lambda_a^{kj} =-C_\rho \delta_i^j$ and $\psi^2$
is the square of  the longest root. (We adopt a convention in which the
shortest root squared is 2 for all the Lie algebras). The central extension in
the affine Lie algebra is parametrized such that the unitary highest weight
representations exist for positive integer values of $k$. $C_v$ is the
eigenvalue of the second Casimir in the adjoint representation of $g$ related
to the dual Coxeter number $h_g^\vee$ by $C_v=\psi^2 h_g^\vee$.
The raising and lowering of the indices $i,j,..$ is done by the
metric $\eta_{ij}=-\epsilon
\eta_{ji}$, satisfying the relation $\eta_{ik}\eta^{jk}=\delta_i^j$, by the
rule: $V^i=\eta^{ij}V_j$ and $V_i=V^j \eta_{ji}$, for any quantity $V$.
The parameter $\epsilon=-1$ for the superconformal algebras and $\epsilon=+1$
for the quasisuperconformal algebras. For $\epsilon=-1$ the currents
$G^i(z)$ are fermionic, while for   $\epsilon=+1$ they are bosonic.

The tensor $P^{ab}_{ij}$ is defined by
$$
P^{ab}_{ij}=  \lambda^a_{ik}\lambda^{bk}_{~~j}+
          \lambda^b_{ik}\lambda^{ak}_{~~j}+
{2\over \sigma_0} \eta_{ji}\delta^{ab}\ . \eqno(2.4)
$$
Note the symmetry properties: $P^{ab}_{ij}=-\epsilon P^{ab}_{ji}$ and
$\lambda^a_{ij}=\epsilon \lambda^a_{ji}$. Thus
$\eta^{ij}\lambda^a_{ij}=0$. The OPE algebra (2.1) closes
provided that the parameters occurring in the algebra obey the following
relations [4]
$$
\eqalignno{
\gamma =& {d(d+\epsilon)\over \psi^4 D i_\rho\eta }\ , \qquad\qquad
\sigma = {2d\over \psi^2 D i_\rho\eta}[(d+\epsilon)(k+h_g^\vee)-Di_\rho] \ ,
 &(2.5a,b)\cr
 c=&3k\psi^2\sigma +{k\over \eta}(D+\epsilon d +1)\ ,\qquad\qquad
 b={k\psi^2\sigma\over 2} \ , &(2.5c,d)\cr}
$$
where
$$
\eta = k+h_g^\vee+\epsilon i_\rho\ . \eqno(2.6)
$$
We now turn to the construction of the BRST operator corresponding to the
above algebra. We introduce the pairs of ghosts $(b,c)$, $(\beta^i,\gamma_i)$
and $(r^a, s_a)$, corresponding to the generators $T, G^i$ and $J^a$,
respectively. The ghosts $(c,\gamma^i, s^a)$ have ghost number $1$ and
conformal dimension $(-1,-\ft12, 0)$, respectively, while the antighosts
$(b,\beta_i,r_a)$ have ghost number $-1$ and conformal dimension $(2,\ft32,
1)$, respectively. They satisfy the following OPEs:
$ c(z)b(\omega)=(z-\omega)^{-1}+\cdots$,
$\gamma^i(z)\beta_j(\omega)=\delta^i_j (z-\omega)^{-1}+\cdots$
and $s^a(z)r^b(\omega)=\delta^{ab}(z-\omega)^{-1}+\cdots$.

Using the result of
[10] we make an ansatz for the BRST operator depending on a number of
parameters. We then verify fully the nilpotency of the BRST operator, which
fixes all these parameters and in addition  imposes conditions on
the parameters of the algebra (2.1).  For the BRST operator we find the
following result:
$$
\eqalign{
  Q=& cT+\gamma^iG_i+s^a J_a+bc\partial c+\beta_i\big(\ft12 \gamma^i\partial
c-\partial\gamma^i c\big)\cr
 &-r_as^a\partial c-b\gamma_i\gamma^i +\lambda_{ai}^{~~j}\big( -\ft12
\epsilon\sigma_0 r^a\gamma^i\partial\gamma_j+s^a\beta^i\gamma_j\big)
-\ft12 f_{ab}^{~~c} r_cs^as^b\cr
& -\ft12 \gamma P^{ab}_{ij}J_ar_b\gamma^i\gamma^j-\ft1{24}\gamma^2 P^{ab}_{ij}
P^{cd}_{kl}f_{ac}^{~~e}
r_b r_d r_e\gamma^i\gamma^j\gamma^k\gamma^\ell \ .\cr}\eqno(2.7)
$$
Using the relations (2.5a,b) and various group theoretical relations provided
in [5], we find that the above BRST operator is nilpotent provided that
the central extensions satisfy the following relations:
$$
k=-2(h_g^\vee+\epsilon i_\rho)\ ,\quad\quad
c= 26+11\epsilon d+2D\ ,\quad\quad b= 16+6\epsilon d \ .\eqno(2.8a,b,c)
$$
In (2.8b) the central charge equals the sum of contributions
$2(-1)^{2s}(6s^2-6s+1)$ from each generator of conformal dimension $s$ with  an
additional factor of $-\epsilon$ for spin 3/2 generator. The relations
(2.8b,c) agree with (2.5c,d),
upon the use of (2.8a) and particular values of various group theoretical
quantities
listed in the Table. The crucial new information
implied by the existence of
the quantum BRST operator is the condition  (2.8a) on the affine Lie
algebra level $k$. From the Table we see that $k$ will always be negative (for
$N>4$). This is potentially a problem in obtaining unitary representations of
the algebra.

For the special case of $SO(N)$--extended superconformal algebras, the level
$k$ and the Virasoro central extension $c$ are [10]
$$
k=6-2N\ , \qquad\qquad c=26-12N+N^2\ . \eqno(2.9)
$$
Note that once we restrict ourselves to the region
$N>4$, we have $c>0$ for $N\ge 10$. Furthermore, $c<0$ for the cases $g=B_3$
and $g=G_2$. It should be emphasized, however, that so far we have considered
the complex form of the Lie algebra $g$. In applications, it is usful to
consider its real forms. For example, for $D_n$, the real forms are $so(2n)$,
$so(p,q),\ (p=q=2n)$ and $so^*(2n)$. Using the noncompact real forms, and
imposing an invariance condition under their maximal compact subalgebras, one
may in principle obtain unitary representations. The implications of such a
construction remains to be seen.

In ref. [9], we also found the conditions for the existence of nilpotent BRST
operator for quadratically nonlinear superconformal algebras whose
current algebra sector is based on $osp(N|2M)$ or $s\ell(N+2|N)$. In
the former case, the conditions on $k$ and $c$ are exactly of the form given in
(2.9) but with $N$ replaced by the superdimension $d_s \equiv N-2M$, while in
the latter case we found that $k=-4$ and $c=-12$.
\bigskip
\noindent{\bf 3. Free Field Realisation of Quadratically Nonlinear
Superconformal  Algebras }
\bigskip
The first free field realisation of a quadratically nonlinear algebra to be
constructed was that of $SO(N)$--extended superconformal algebra in terms of
$N$ real fermions $\psi_i(z)$  and a real boson $\phi(z)$ [13].  This
realization has level $k=1$ and therefore is not suitable for the BRST
operator. This realisation was later generalized for arbitrary level $k$ for
the case of $SO(N)$ and $SU(N)$--extended superconformal algebras in ref.
[11] and to all the algebras listed in table in ref. [14]. For a physical
application the $SO(N)$--extended algebra seems to be the most promising at
present, and therefore here we shall consider the realisation of this
particular case only.

The $SO(N)$--extended superconformal algebra is realised in terms of a real
scalar $\phi$ with a background charge $\alpha$, $N$ real fermions $\psi^i$ and
level $\ell$ affine Lie algebra currents $K^a$. The generators of the algebra
are realised in terms of these fields as follows [11]
$$
\eqalign{
T=&  -\alpha \partial ^2 \phi - \ft12 \partial \phi \partial\phi
-\ft12 \psi\partial \psi -{1\over 2\eta} K^aK_a, \cr
G_i=& 2i\alpha \partial\psi
+ i\partial \phi \psi - 2i \alpha \ell^{-1} \lambda^a_{ij} K_a \psi^j, \cr
J_a=& K_a+ \ft12 \lambda^a _{ij} \psi^i \psi^j\ .\cr}  \eqno(3.1)
$$
These generators obey the algebra (2.1) with $b,c,\gamma, \sigma$ given in
(2.5) evaluated for $SO(N)$ and provided that the following additional
relations hold [11]
$$
k=\ell+1\ ,\qquad\qquad \alpha^2={\ell^2\over 4\eta}\ . \eqno(3.2)
$$
The valuse of the central charge can be written a way which
illustrates the contribution of different terms in the energy momemntum tensor
as follows
$$
    c=1+12\alpha^2+{N\over 2}+ {\ell D\over \eta}\ . \eqno(3.3)
$$

The existence of the quantum BRST operator imposes the restriction $k=6-2N$
 which in turn restricts the value of the background charge to be
$$
     \alpha = {2N-5\over 2\sqrt{3-N}}\ . \eqno(3.4)
$$
Note that for $N>4$ the background charge becomes imaginary. The
free field realisation of the level $\ell$ currents $K^a$ can
be achieved by the Wakimoto construction [15] which uses $N/2$ free scalars
and additional commuting fermionic fields.
	The $N/2$ free scalars do not seem to lend themselves
to a spacetime interpretation. The real challenge is to find a multi-scalar
realisation which will allow such an interpretation.
\bigskip
{\bf 4. A Nonlinear Superconformal Algebra Based on $S^7$ }
\bigskip
	A characteristic feature of the algebras that we have considered so
far is that the OPE of the two spin 3/2 currents contains an operator bilinear
in spin 1 currents. In this section we will discuss an $N=8$ nonlinear
superconformal algebra based on seven-sphere $S^7$ [12] which is
a  different type of nonlinear algebra. It is a soft
algebra because  its structure `constants' are not constants but depend
 on a point of the seven-sphere. A certain version of this
algebra  arises in a twistor-like formulation of $D=10$ dimensional
Green-Schwarz superstring as a symmetry on the world-sheet [16], and the
current algebra sector was constructed sometime ago [17]. The algebra
constructed in ref. [12] contains the energy-momentum tensor $T$, fermionic
dimension 3/2 operators $G^a$ , $(a=0,1,...,7)$, and a multiplet of spin 1
currents $J^i$ $(i=1,2,...,7)$. The structure functions of the algebra depend
on the coordinates of seven sphere parametrized by unit length octonions X.
Namely, $S^7=\{ X\in {\bf O} |\ |X|=1 \}$. We can express $X$ as
$X=X^ae_a=X^0e_0+X^ie_i$ where the octonionic units are $e_a=(1,e_i)$ and
$e_i$ obey the  algebra $e_ie_j=-\delta_{ij}+c_{ijk} e_k$ where $c_{ijk}$ are
the octonionic structure constants. Octonionic conjugation is defined as
$X^*=X^0e_0-X^ie_i$.  Using the notation of ref. [12], we define $[X]=\ft12
(X+X^*)$ and  $\{X\}=\ft12 (X-X^*)$.

The  construction of the algebra of ref. [12] proceeds by setting
$X=\lambda/ |\lambda |$ and introducing the real scalar fields
$\phi^I\ (I=0,...,7)$ , the anticommuting fermions $S^a$, bosons
$(\lambda^a,\omega^a)$ with conformal weights (1/2,1/2), and their
anticommuting superpartners $(\theta^a,\pi^a)$ of conformal weights (0,1).
(While  $\phi^I$ and $S^a$ are candidates for the bosonic and fermionic
coordinates of a target superspace, the interpretation of the other
fields is less clear. They may arise in a twistor-like formulation of
superstring theory).  In terms of these fields the generators of the algebra
are represented as follows [12]
$$
\eqalign{
J^i=& -\Gamma^i_{ab}(1)\omega^a\lambda^b -\ft12 \Gamma^i_{ab}(X)S^aS^b\ ,\cr
G_a=& c_{abc}(1)\big( \pi^b\lambda^c-\partial\theta^b\omega^c\big)
+\lambda^I_{ab}(X)\partial\phi^I S^b
       -t_{abcd}(X)|\lambda|^{-1}\partial\theta^bS^cS^d\ ,\cr
 T=&\ft12\big(\partial\lambda^a\omega_a-\lambda^a\partial\omega_a\big)
-\pi^a\partial\theta_a+\ft12\partial\phi^I\partial\phi^I
 -\ft12 S^a\partial S_a\ ,\cr}\eqno(4.1)
$$
where
$$
\eqalign{
        \Gamma^i_{ab}(X)=& \bigg[\big((Xe_a)^*(Xe_b)\big)e_i\bigg]\ ,\quad
\quad
   \lambda^I_{ab}(X)=\bigg[\big((Xe^I)^*(Xe_b)\big)e_a^*\bigg]\ , \cr
c_{abc}(X)=&\bigg[\big((Xe_b)^*(Xe_c)\big) e_a^*\bigg]\ , \quad\quad
t_{abcd}(X)=(\delta^{fb}-X^fX^b)
\bigg[\big\{(Xe_c)^*(e_fe_d)\big\}e_a^*\bigg]\ . \cr}\eqno(4.2)
$$

Using the two point functions implied by the form of the energy-momentum
tensor, and keeping only the single contraction terms, which amounts to the
calculation of the {\it classical} algebra, for the nontrivial part of this
 algebra one finds [12]
$$
\eqalign{
J_i(z)J_j(\omega) =& {2 \over (z-\omega)}c_{ijk}(X) J_k+\cdots \cr
J_i(z)G_b(\omega) =& {1 \over (z-\omega)}
\biggl(\tau^i_{ab}(X) G^b(\omega)
+\bigl(\lambda_{ijabc}-\tau_{ijabc}(X)\bigr)
       |\lambda|^{-2}\lambda^b\partial\theta^c J^j\biggr)+\cdots \ , \cr
G_a(z)G_b(\omega) =& -{2 \over (z-\omega)^2}\Gamma^i_{ab}(X)J_i(\omega)
-{1\over (z-\omega)} \biggl(\Gamma^i_{ab}(X) \partial J_i(\omega)
+ 2\delta_{ab} T(\omega) \biggr)+\cdots  \cr }\eqno(4.3)
$$
where
$$
\eqalign{
\tau^i_{ab}(X)=&\biggl[\bigl((e_aX^*)(Xe_i)\bigr)e_b^*\biggr]\ ,\cr
\lambda_{ijabc}=&\biggl[\bigl(e_b^*\bigl((e_ce_a)e_i)\bigr)e_j\biggr]\ ,\cr
\tau_{ijabc}(X)=&\biggl[\bigl(e_b^*(e_c((e_aX^*)(Xe_i)))\bigr)e_j\biggr]\ .\cr}
\eqno(4.4)
$$

The quantum version of this algebra is not known. However, the
quantum version of the $S^7$ loop algebra and the Sugawara
construction based on it has been discussed in [18]. A classical BRST operator
associated with the $S^7$ loop algebra is also given in ref. [18].
\bigskip
\noindent{\bf 5. Conclusions}
\bigskip
If the quadratically nonlinear superconformal algebras are to find
applications in the construction of a new type of string theory, there are two
main obstacles that need to be overcome. Firstly, a multi-scalar realisation
that will allow a spacetime interpretation needs to be found. Secondly,
the issue of unitarity needs to be addressed since the existence of
quantum BRST operator fixes the level of the affine Lie algebra sector to be
negative. Taking the affine Lie algebra sector to be based on a noncompact Lie
algebra and then imposing an invariance condition under its maximal compact
subalgebra may provide a solution. Whether this is indeed the case and the
implications of such a construction as far as spacetime interpretation is
concerned remains to be seen.

Finally, an alternative way of going beyond local $N=4$ worldsheet
supersymmetry in string theory may be based on far more nonlinear extended
superconformal algebras, one of which is due to ref. [12], as described in
the last section. Another example may be provided by the $N=8$ conformal
supergravity of ref. [19].

\np

\centerline{\bf References}
\bigskip
\item{1.} M. Ademollo et al., Phys. Lett. {\bf B208} (1976) 447; Nucl. Phys.
{\bf B111} (1976) 77.
\item{2.} V.G. Knizhnik, Theor. Math. Phys. {\bf 66} (1986) 68.
\item{3.} M. Bershadsky, Phys. Lett. {\bf B174} (1986) 285.
\item{4.} E.S. Fradkin and V. Y. Linetsky, Phys. Lett. {\bf B282} (1992)
         352;{\bf B291} (1992) 71; Phys. Lett. {\bf B275} (1992) 345.
 \item{5.} P. Bowcock, Nucl. Phys. {\bf B381} (1992) 415.
\item{6.} L.J. Romans, Nucl. Phys. {\bf B357} (1991) 549.
\item{7.} F.A. Bais, T. Tjin and P. van Driel, Nucl. Phys. {\bf B357}
          (1991) 632;
\item{} A.M. Polyakov, in ``Physics and Mathematics of Strings'' (World
             Scientific, 1990);
\item{} M. Bershadsky, Commun. MAth. Phys. {\bf 139} (1991) 71.
\item{8.} F. Defever, W. Troost and Z. Hasiewicz, Phys. Lett. {\bf B273}
(1991) 51.
\item{9.} Z. Khviengia and E. Sezgin, {\it BRST operator for superconfomal
algebras with quadratic nonlinearity}, preprint, CTP TAMU-27/93.
\item{10.} K. Schoutens, A. Sevrin and P. van Nieuwenhuizen, Commun. Math.
           Phys. {\bf 124} (1989) 87.
\item{11.} P. Matthieu, Phys. Lett. {\bf B218} (1989) 185.
\item{12.} L. Brink, M. Cederwall and C.R. Preitschopf, {\it N=8
  superconformal algebras and the superstring}, preprint, G\"oteborg-ITP-93-5.
\item{13.} K. Schoutens, Nucl. Phys. {\bf B314} (1989) 519.
\item{14.} K. Ito, J.O. Madsen and J.L. Petersen, in Proceedings of
the International Workshop on {\it String Theory, Quantum Gravity and
the Unification of Fundamental Interactions}, Rome, 1992.
\item{15.} M. Wakimoto, Commun. Math. Phys. {\bf 104} (1986) 605;
\item{} K. Ito and S. Komata, Mod. Phys. Lett. {\bf A6} (1991) 581.
\item{16.} N. Berkovits, Nucl. Phys. {\bf B358} (1991) 169.
\item{17.} F. Englert, A. Sevrin, W. Troost, A. van Proeyen
            and  P. Spindel, J. Math. Phys. {\bf 29} (1988) 281.
\item{18.} M. Cederwall and C.R. Preitschopf, {\it $S^7$ and ${\hat {S^7}}$},
preprint, G\"oteborg-ITP-93-34.
\item{19.} E. Bergshoeff, H. Nishino and E. Sezgin, Phys. Lett. {\bf B218}
           (1987) 167.

 \end